\documentclass[manuscript]{aastex}
\usepackage{multirow}

\begin{document}
\title{Population synthesis on high-mass X-ray binaries: prospects and constraints from the universal X-ray luminosity function}
\author{Zhao-Yu Zuo$^{1,3}$, Xiang-Dong Li$^{2,3}$ and Qiu-Sheng Gu$^{2,3}$}
\affil{$^1$Department of Physics, School of Science, Xi'an Jiaotong University, Xi'an 710049, China\\
$^2$Department of Astronomy, Nanjing University, Nanjing 210093, China;\\
$^3$Key laboratory of Modern Astronomy and Astrophysics (Nanjing
University), Ministry of Education, Nanjing 210093, China
\\zuozyu@mail.xjtu.edu.cn; lixd@nju.edu.cn}

\begin{abstract}

Using an updated population synthesis code initially developed by Hurley et al. we modeled
the synthetic X-ray binary (XRB) populations for direct comparison with
the universal, featureless X-ray luminosity function (XLF) of high mass X-ray
binaries (HMXBs) in star-forming galaxies. Our main goal is to use the
universal XLF to constrain the model parameters, given the current knowledge
of binary evolution. We find that the one-dimensional (1D) Maxwellian velocity dispersion of the natal kick
can be constrained to be of the order of $\sigma_{\rm kick}\sim 150\,\rm km\,s^{-1}$, supporting
earlier findings that neutron stars (NSs) formed in binaries seem to receive significantly smaller natal kicks than
the velocities of Galactic single pulsars would indicate. The super-Eddington accretion
factor is further confirmed in the framework of stellar mass black holes (BHs), revealing the true origin of the most of the
ultraluminous X-ray sources (ULXs) may indeed be the high-luminosity extension of
ordinary HMXBs which harbor stellar-mass BHs rather than exotic intermediate-mass
BHs or ones. We present the detail properties of the model-predicted present-day HMXBs, which may be investigated by future
high-resolution X-ray and optical observations of sources in nearby star-forming galaxies.

\end{abstract}


\section{Introduction}

High mass X-ray binaries (HMXBs) are binary systems, in which a high mass primary
star formed the compact star accreting from a secondary massive star.
They are conventionally divided into two subgroups \citep{paradijs83}.
One group usually contains an evolved (super)giant star, generally $M_{\rm opt}\ga 10M_{\odot}$, having strong
stellar wind or filling its Roche lobe (RL) to power a  bright X-ray source for $\sim
10^{5}-10^{6}$ yr. The compact star should be either a neutron star (NS;
$M\sim 1.4 M_{\odot}$) or a stellar-mass black hole \citep[BH; $\thicksim 3<M/M_{\odot}< \sim20$ for
solar metallicity, but may reach $\sim 100$ in metal-poor environments, see][for reviews]{rm06}
as a result of collapse of high mass primary star. Another possible type of accreting
objects may be intermediate-mass ($\sim 10^2\leq M/M_{\odot}\leq 10^5$) BH \citep[i.e., IMBHs, see][for a review]{marel04},
however its exact origin is still not well understood.
The other group is so-called Be/X-ray binaries (Be-XRBs), in which it contains a
Be (B-type star which shows emission-line spectra) companion, usually
accreted by an NS during its periastron passage, showing as X-ray transients.

With Chandra's unprecedented sensitivity and angular resolution \citep{Weisskopf00},
a large number of HMXBs have been discovered in galaxies even beyond the Local Group \citep{fabbiano06},
allowing to do studies of the collective properties of HMXB populations as a whole.
One of the most striking features of HMXB populations is that the X-ray luminosity function (XLF)
takes a possibly universal form of a single, smooth power law giving an excellent account of HMXBs containing
NSs, stellar-mass BHs and probably IMBHs over the entire X-ray luminosity range
$\sim 10^{35} \leq L_{\rm X} \leq 10^{40} \rm ergs\,s^{-1}$. This is first discovered by \citet{ggs03},
based on {\it Chandra\/} and {\it ASCA\/} data of nearby star-forming galaxies and
{\it RXTE/ASM\/}, {\it ASCA\/}, and {\it MIR-KVANT/TTM\/} observations of
our Galaxy and the Magellanic Clouds. They showed that for a wide range of star formation rate (SFR),
the HMXB XLF in a galaxy can be well described by a power law with slope of $\sim$1.6, the normalization of which is proportional
to the SFR. They searched for but found no features corresponding to the Eddington luminosities of NS and BH
in the averaged XLF. They argued, however, that the expected features may
be smeared and diluted by various effects such as distance uncertainties.
With a larger sample of galaxies and better control of systematic effects, \citet[][hereafter MGS for short]{mineo12} revisited
this problem and found that the average HMXB XLF is entirely consistent with the one
obtained by \citet{ggs03}. The accuracy of XLF slope has been improved to $\sim0.03$, and
the values of the high luminosity break at $L_{\rm X} \sim 10^{40} \rm ergs\,s^{-1}$ are consistent
within statistical uncertainties. They did not find any statistically
significant feature in the XLF near the critical Eddington luminosity
of NSs, either.

Although the absence of features in the HMXB XLF is striking and puzzling, theoretical
investigations on this remain limited. Using the fundamental mass-luminosity and mass-radius
relations for massive stars, as well as a natural assumption on the power-law initial
mass function (IMF; Salpeter IMF or Miller-Scalo IMF) and following a semi-empirical approach,
\citet{postnov03} noted that the universal XLF can be readily explained by the universal
properties of mass transfer rates in HMXBs. \citet{bl08}, using the ``Scenario Machine''
code \citep{lipunov09}, instead argued that there should be no universal XLF in both observational
and theoretical aspects. They suggested that the evolution of binaries and their lifetimes
in their X-ray stages should be taken into account in future theoretical modelings.
Recently, \citet{bg12} used a Jacobian transformation method
to calculate the XLF and the binary-period distribution of HMXBs in the stellar fields of
normal galaxies. Their model XLF can match the observed XLF shape quite closely.
They suggested that a future Monte Carlo evolutionary population synthesis (EPS) scheme is promising to obtain
more detailed understanding of the formation and evolution of HMXB populations.

In fact population synthesis studies on the XLF have already been examined and explored extensively in
the past decade. Several authors focused on the XLF modeling for individual galaxies, the types of 
which cover almost the entire galaxy morphological sequence, for example the star forming galaxies 
\citep[see][i.e., NGC 1569]{bel04},  star-burst galaxies \citep[][NGC 4038/4039, the Antennae]{liu07}, 
and elliptical galaxies \citep[][NGC 3379 and NGC 4278]{f08,f09}. Specifically \citet{Linden10,Linden11} 
modeled the XLF for HMXBs and Be-XRBs in the Small Magellanic Cloud (SMC). \citet{luo12} studied the 
XLFs of XRB populations in NGC 1291, in both the bulge and ring regions. Additionally \citet{lu12} calculated 
the numbers and birthrates of symbiotic XRBs in the Galaxy. 
Several authors focused on X-ray/XLF evolution or their numbers of a specific type of galaxies  
globally, instead. However most of them are based on semi-empirical, semi-analytical 
approaches, with simplified assumptions adopted on the formation and
evolution of XRBs \citep{wg98,bever00,gw01,wu01,pb02,bd04,rpkr11,bg12,bg13a,bg13b}. 
It is worth noting that the more sophisticated, state-of-the-art EPS simulation has also been explored
in this direction. The first attempt was done recently by \citet{zuo11} on the cosmic X-ray evolution of XRBs in late-type
galaxies, and its dependence on the physical properties of galaxies (e.g., optical luminosity,
stellar mass, and mass-to-light ratio), which is followed by \citet{f13} focusing mainly on the global scaling of
emission from XRB populations with star-formation rate and stellar mass, and their evolution with redshifts, 
by using the Millennium-II simulation as initial conditions. As a series of works following \citet{f13}, 
another two EPS studies are presented recently. One is by \citet{Tremmel12}  studying on the redshift evolution 
of the normal galaxy XLF as well as integrated XRB emission from entire galaxies, the other is by 
\citet{Tzanavaris13} focusing on modeling the XLFs in galaxies in the Spitzer Infrared Nearby Galaxy Survey (SINGS).

In the present work, we use a most up-to-date EPS code to model the observed XLF (both the shape and the absolute source number)
of HMXBs in star-forming galaxies. We also evaluate the effects of several input parameters
such as the IMF of binary stars, the natal kick distribution, common envelop (CE) efficiency and super-Eddington
factor (see Sec~2.1 for details) on the results. One particular objective of this study is to
use the universal featureless XLF to constrain the model parameters. We also aim to explore
the detailed components of HMXB populations, which may help understand the nature of the sources
and may be testified by future observations.

This paper is organized as follows. In \S 2 we describe the
population synthesis method and the input physics for X-ray binaries (XRBs) in our
model. The calculated results and discussion are presented in \S 3. Our conclusions are in \S 4.

\section{Models}

\subsection{Assumptions and input parameters}
We calculate the expected numbers for various types of HMXB population
using a version of EPS code developed by \citet{Hurley00,Hurley02} and
updated as described in \citet[][see Appendix A in their paper]{liu07} and
\citet{zuo08}. In the present code, 
the compact object masses are calculated in a different way than originally suggested
by \citet{Hurley00} and \citet{liu07}. We use a prescription the same as
\citet[][i.e., the Rapid supernova mechanism]{fryer12}, which can reproduce successfully
the mass gap observed in Galactic XRBs when combined with binary evolution \citep{bel12}.
We also allow for the formation of low mass NSs through ECS \citep[][]{Podsiadlowski04}.
The maximum NS mass is assumed to be $2.5\,M_{\odot}$, above which BH is assumed to form.
We change the recipes for mass loss of stellar winds by using the metal-dependent
fitting formulae given by \citet[][see also Belczynski et al. 2010]{vink01}.
The wind velocity is
difficult to determine accurately, and usually set to be proportional to the escape
velocity from the surface of the mass-losing star, as a ratio $\beta_{\rm wind}$.
The values of $\beta_{\rm wind}$ must depend on the spectral type of the mass-losing
star \citep{lamers95,kucinskas99}. We adopt $\beta_{\rm wind}=0.125$ (i.e., slow winds)
for He-rich stars and extended ($R_{\rm don}> 900R_{\odot}$) H-rich giants,
$\beta_{\rm wind}=0.8$ for high-mass ($>1.4 M_{\odot}$) main sequence (MS) stars, and the default value 0.5 for others.

Another two major updates are related to the CE evolution. One (and of the
most important improvements recently) is on the CE coefficient, $\lambda$, which describes
the binding energy of the envelope. We now use a more physical estimate of $\lambda$
\citep[][see below]{xu10,loveridge11} rather than the constant value conventionally
assumed by most previous studies. The other is on the updated critical mass ratio criterion for CE
initiated by Hertzsprung gap (HG) donor stars, recently developed by \citet[][private communication, see Appendix A]{shao12}.
The values of other parameters are adopted the same as the default ones in \citet{Hurley02}
if not mentioned otherwise.

The HMXBs studied by MGS all reside in nearby star-forming
galaxies (see their Table~1 for details). Due to heterogeneous data of these galaxies,
the metallicity estimation for each galaxy is still not available, however the rough
value is most likely to be around subsolar as a whole (private communication with Mineo S.). So we adopted
a fixed subsolar metallicity (0.5$Z_{\odot}$) in our basic model. A lower metal
abundance mainly affect stellar wind, making it weaker, so we designed a ``WEAK" wind model (i.e.,model M8) to test this
effect. For star formation, a constant SFR of 50 Myr is assumed in our basic model.
In each model, we evolve $10^6$ primordial
systems\footnote{We also vary the number  of the binary systems
by a factor of eight, and found no significant difference in the
final results.}, all of which are initially binary systems. We set
up the same grid of initial parameters (primary mass, secondary
mass and orbital separation) as \citet{Hurley02} did and then evolve
each binary. In the following we describe the
assumptions and input parameters in our basic model (i.e., model
M1, listed in Table 1).

\noindent {\em (1) initial parameters}\\
We take the IMF of \citet[][hereafter KROUPA01, with power law slope of -1.3
in 0.08-0.5$M_{\odot}$, and -2.3 in 0.5-80.0$M_{\odot}$]{Kroupa01} for the
distribution of the primary mass ($M_1$). For the secondary's mass ($M_2$),
a uniform distribution is assumed for the mass ratio $M_2/M_1$ between 0 and 1.
We also adopt a uniform distribution for the logarithm of the orbital separation $\ln a$
\citep{Hurley02}.

\noindent {\em (2) CE evolution} \\
When mass transfer becomes dynamically unstable, a binary may enter a CE phase.
An important parameter determining the outcome of the CE is the CE parameter $\alpha_{\rm CE}$
\citep{paczynski76,iben93}. It describes the efficiency of converting orbital energy
(\textbf{$E_{\rm orb}$}) into the kinetic energy,
resulting in the ejection of the envelope ($M_{\rm env}$).
We use the standard energy prescription \citep{webbink84,Kiel06} to compute the
outcome of the CE phase.
\begin{equation}
E_{\rm bind}=
\alpha_{\rm CE}[\frac{GM_{\rm c}M_{2}}{2 a_{\rm f}}-\frac{GM_{\rm
c}M_{2}}{2 a_{\rm i}}],
\end{equation}
where $G$ is the gravitational constant, $a_i$ and $a_f$ denote the initial and final orbital separations, respectively;
$M_{\rm c}$ is the helium-core mass of the primary star ($M_1$); $E_{\rm bind}$
the binding energy of the hydrogen-rich envelope. Conventionally, 
a so-called envelope-structure parameter, $\lambda$, defined by
\begin{equation}
E_{\rm bind}=-\frac{GM_1M_{env}}{R_{L_1}\lambda}
\end{equation}
is used to compute the binding energy, where $R_{L_1}$ is the RL radius of the primary star. The parameter $\lambda$
is often assumed to be a constant value \citep{Hurley02,zuo10}, however in reality it will
vary for stars of different masses and different evolutionary phases, far from constant \citep{dewi00,sluys06}.
Recent work by \citet{loveridge11} presents accurate analytic prescriptions of the envelope binding energy for giants
as a function of basic stellar parameters such as the metallicity, mass, radius, and evolutionary phase of the star.
They computed the envelope binding energy $E_{\rm bind}$ by integrating the gravitational and internal energies from the core-envelope
boundary to the surface of the star $M_s$ as follows,
\begin{equation}
E_{\rm bind}=\int^{M_s}_{M_c}(E_{in}-\frac{Gm}{r(m)})dm,
\end{equation}
where $E_{in}$ is the internal energy per unit of mass, containing terms such as
the thermal energy of the gas and the radiation energy, but not the recombination
energy \citep[for more details, see][]{sluys06}. Here we adopt \citet{loveridge11}'s prescription for CE evolution.

\noindent {\em (3) super-Eddington radiation}\\
In the literature it is often implicitly assumed that the luminosities of
accreting NS/BH binaries were constrained by the critical Eddington limit:
\begin{equation}
L_{\rm X} \lesssim L_{\rm Edd} \simeq
\frac{4\pi GM_{1}m_{\rm p}c}{\sigma_{\rm T}}=1.3 \times
10^{38}\frac{M_{1}}{M_{\sun}} \,ergs\,s^{-1},
\end{equation}
where $\sigma_{\rm T}$ is the Thomson cross section, $m_{\rm p}$ is the proton mass and
$c$ the velocity of light. However we note that in reality
this limit may fail for several systems. One possible example is the recently
discovered large population of ultra-luminous X-ray
sources (ULXs, non-nuclear point-like sources with isotropic X-ray luminosity
exceeding $10^{39} \rm ergs\,s^{-1}$), often associated with star-forming regions \citep{zezas99,rw00,fzm01}.
Its luminosity is even higher than the Eddington luminosity for a $\thicksim 10 M_{\odot}$ accreting BH.
Several different scenarios have been proposed to explain its origin, such as HMXBs powered by stellar mass
BHs with anisotropic X-ray emission \citep[``beaming'' model,][]{king01} or with super-Eddington accretion rate/luminosity
\citep[due to photon bubble instability,][]{Begelman02} or with a combination of the two mechanisms \citep{king08}.
Accretion binaries with a BH mass of $10^2M_{\odot}<m_{\rm BH}<10^5M_{\odot}$ (IMBHs) can also be a possibility, however can hardly account for all the ULXs observed in the galaxies, but only
the most luminous sources (with $L_{\rm X}\gtrsim10^{41} \rm \,ergs\,s^{-1}$). Here we introduce a
parameter $\eta_{\rm Edd}$, i.e. ``Begelman factor" \citep{rappaport04} to examine the
allowed maximum super-Eddington accretion rate if powered by stellar mass BHs.
 In our basic model, $\eta_{\rm Edd}$ is adopted as 100 for BH XRBs. We
also reduce its value to 80, 50, 30, 10 and 5 (i.e., model M4),  to examine its effect.
On the other hand, NS accretors seem to provide at most several times the Eddington-limited luminosity \citep{Levine91,Levine93,grimm03,rpp05,f08},
here we adopt $\eta_{\rm Edd}$ for NS XRBs as 5 and keep this assumption throughout.

\noindent {\em (4) SN kicks}\\
At the time of birth NSs and BHs receive a velocity kick due to any asymmetry in the supernova (SN)
explosions \citep{lyne94}. We assign a Maxwellian kick distribution with
a dispersion velocity of $\sigma_{\rm kick}=150\, \rm km\,s^{-1}$ for newborn NSs in our basic model. 
For BHs, we scale down the natal kick by multiplying the kick by the fraction
of material which does not fall back onto the compact objects. Additionally, BHs
formed with small amounts of fall back ($M_{\rm fb}<0.2M_{\odot}$) are assumed to
receive full kicks. In situations where BHs form silently (without a SN explosion) via direct collapse, we apply no
natal kick in our basic model \citep{fryer12,Dominik12}. Moreover, for ECS NSs,
no natal kick is assumed since these are weak SN occurring for the lowest stars \citep[$M_{\rm ZAMS}=7.6 - 8.3 M_{\odot}$,][]{Hurley00,et04a,et04b,bel08}.

We also construct several other models by varying the key
input parameters, as listed in Table~1.

(1) Variations of the CE parameter can change the orbital separation
of the binary considerably, resulting in different outcomes of the final evolution.
However a reliable value for $\alpha_{\rm CE}$ is difficult to estimate
due to a lack of understanding of the complicated processes involved, although
it is adopted extensively from $\sim$ 0.1 to $\sim$
3.0 (e.g., Taam \& Bodenheimer 1989; Tutukov \& Yungelon 1993;
Podsiadlowski, Rappaport \& Han 2003) in the literature. Here we adopt
$\alpha_{\rm CE}=0.5$ in our basic model and change it to 1.0 (model M2) to examine its effect.

(2) Surveys of M dwarfs within 20 pc from the Sun have indicated that the binary
fraction $f$ may be a function of stellar spectral types \citep{fischer92}.
For example, recent works by \citet{lada06} and \citet{kobulnicky07} find that
for G stars $f>0.5$ while $f>0.6$ for massive O/B stars in the Cygnus OB2 association.
So we set $f=0.5$ in our basic model and modify it to $f=0.8$ (model M3) for comparison.

(3) Observations show that compact young massive clusters contain
more massive stars preferentially \citep{sternberg98,smith01}, so we also
make use of the IMF of \citet[][hereafter MT87, with power law slope of -1.3
in 0.08-1.0$M_{\odot}$, but -1.95 in 1.0-80.0$M_{\odot}$, model M6]{mt87}, which is
more skewed towards high mass than in KROUPA01. For the mass of the secondary star ($M_2$), a power-law
distribution of $P(q)\propto q^{\alpha}$ is assumed, where $q\equiv M_2/M_1$.
We adopt both the conventional choice of flat mass spectrum, i.e., $\alpha=0$
\citep[our basic model, M1,][]{Mazeh92,goldberg94,Shatsky02} and $\alpha=1$ (model M5),
since recent observations are more in accord with ``twins" being a general feature of the
close binary population \citep{dalton95,kobulnicky07}.

(4) The kick velocity can affect not only the global velocity of the binary system \citep{zuo10}
but also the outcome of the XRB evolution. Though the research on natal SN kicks has already had a
long history \citep{Bailes89}, its functional form of the underlying speed distribution is
still poorly constrained. Measurements of proper motions for isolated radio pulsars indicate
the typical kick speed is in excess of $\sim100-200\, \rm km\,s^{-1}$ \citep{lyne94,hansen97,cc98,acc02,Hobbs05},
however recent observation of NSs found in binaries seems to reveal that they receive a smaller
natal kick \citep{Pfahl02,bel10b,wwk10,btrb12}, of the order of $100\, \rm km\,s^{-1}$.
So we also adopt $\sigma_{\rm kick}=265$ $\rm km\,s^{-1}$ \citep[i.e., model M7,][]{Hobbs05},
190 $\rm km\,s^{-1}$ \citep{hansen97}, 170 $\rm km\,s^{-1}$ \citep{bel10b}, 100 $\rm km\,s^{-1}$,
and 50 $\rm km\,s^{-1}$ \citep{btrb12} for comparison.

(5) Stellar winds from massive stars show a number of puzzles contrasting
observational and theoretical aspects. The two most prominent are the ``wind clumping"
\citep[e.g.,][]{of59,markova04,Repolust04,lm08}
and ``weak wind problem" \citep[e.g.,][]{cg91,kpg91,hpn02}.
The former is related to the fact that mass loss rates might be twice overestimated since stellar
winds might be forming dense clumps rather than being distributed uniformly.
The latter suggests that wind mass loss rates from late O and early B type stars reveal
a severe drop, by a factor of $\sim$100 than theoretically predicted.
Based on this we reduce the wind mass loss rates by a factor of 2 to examine
its effects (e.g., weak winds, model M8). This is done for all stars at all points
in their nuclear evolution.

\subsection{X-ray luminosity and source type}

For super-giant/main-sequence (SG/MS) HMXBs we use the same methods to compute the $0.5-8$ keV X-ray
luminosities and divide types of different sources as in \citet{zuo11}.
Accreting NS/BH in XRBs are powered by either disk fed by RLOF or stellar wind.
When a star expands to fill its RL, a disk may form transferring masses to
the compact star. Otherwise, wind accretion is needed
to power an observable X-ray source. For wind accretion, we explore the
classical \citet{Bondi44}'s formula to calculate the mass transfer rate to
the compact star. In the RLOF case, we discriminate transient
and persistent sources using the criteria of \citet[][i.e., Eq~36 therein]{l01} for MS
and red giant stars, and of \citet[][i.e., Eqs~20 and ~24 therein]{ivanova06} for white dwarf (WD)
donors, respectively. For transient systems, the duty cycle (DC) is empirically
thought to be less than $\sim$1\% \citep{taam00}. We adopt DC=1\% (probability of finding a system in outburst) in our calculations.
The corresponding X-ray luminosity form is as follows:
\begin{eqnarray}
L_{\rm X, 0.5-8 keV}&=&\left\{
\begin{array} { ll}
  \eta_{\rm bol}\eta_{\rm out}L_{\rm Edd}&\ \rm transients\ in\ outbursts, \\
  \eta_{\rm bol}\min(L_{\rm bol},\eta_{\rm Edd}L_{\rm Edd})&\ \rm persistent\
  systems,
\end{array}
\right.
\end{eqnarray}
where the bolometric luminosity $L_{\rm bol}\simeq
0.1\dot{M}_{\rm acc}c^2$ (where $\dot{M}_{\rm acc}$ is the average
mass accretion rate), $\eta_{\rm bol}$ the bolometric correction factor
converting the bolometric luminosity ($L_{\rm bol}$) to the $0.5-8$ keV
X-ray luminosity \citep{bel04}, adopted as 0.4 though its range is
$\sim 0.1-0.5$ for different types of XRB. $\eta_{\rm Edd}$ is
the 'Begelman' factor to allow super-Eddington luminosities, as stated above.
For transient sources the outburst luminosity is taken as a fraction ($\eta_{\rm out}$) of the critical
Eddington luminosity. We take $\eta_{\rm out}=0.1$ for NS transients and $\eta_{\rm out}=1$ for BH transients with
orbital period $P_{\rm orb}$ less and longer than 1 day and 10 hr, respectively \citep{chen97,Garcia03,bel08}.

We adopt a phenomenological way to define Be-XRBs as in \citet{bz09}.  A HMXB is recognized as
Be-XRB if: (1) it hosts NS accretors. We do not consider BH Be-XRBs since no such system
has been found so far \citep{lph05,lph06}; (2) the donor should be massive ($M_{\rm donor} \geq 3.0 M_{\odot}$) MS star (i.e., O/B star with burning H in its core); (3) accretion proceeds only via stellar wind (no RLOF); (4) only systems
with orbital period in the range of 10-300 days are considered; (5) only a fraction $f_{\rm Be}=0.25$
of the above systems are designated as hosting a Be star, as this seems to
be the fraction of Be stars among all regular B-type stars \citep[$\sim 1/5-1/3$,][]{s88,z02,mg05}.
So technically, we randomly selected only 25\% ($f_{\rm Be}=0.25$) of the massive binaries hosting
a B/O star to predict their numbers in our EPS calculations. The X-ray luminosities of Be-XRBs are
estimated based on its orbital periods using Eq.~11 presented by \citet{dll06}, the formula of which is
obtained by fitting the observed data for 36 Be-XRBs compiled by \citet{rp05}.  For Be transients, the outbursts are short-lived,
typically covering a relatively small fraction of the orbital period \citep[$\sim 0.2-0.3 P_{\rm orb}$,][]{reig11}.
Here we adopt an upper value $DC_{\rm max}=0.3$ to give the expected maximum source numbers.

\section{Results and Discussion}

Based on a population of $\sim700$ compact sources, MGS constructed the
average XLF of HMXBs in galaxies. The HMXB XLF they derived follows a power law with a
 slope of 1.6 in the broad luminosity range $log\,L_{\rm X} \thicksim 35-40$ and shows a moderately
 significant evidence for a luminosity break or cut-off at $log\,L_{\rm X} \thickapprox 40$. In
 addition, they did not find any significant features at the Eddington limit for NS or a
 stellar mass BH. Moreover when compared with each individual galaxy in their primary
 sample, which is normalized to their respective SFRs, there are still considerable dispersions in  
 the amplitude (i.e., total number of HMXBs per unit SFR).
 Here we modeled the HMXB XLF from a theoretical point of view. The 
results are presented below.

\subsection{Comparison with \citet{mineo12} and model predictions}

We adopted several models with different assumptions for the input parameters (see Table~1).
Specifically the input parameters in our basic model (i.e.,
model M1) are SFH$=50$ Myr, $\alpha=0$, $\alpha_{\rm CE}=0.5$, $\eta_{\rm Edd}=100$,
$f=0.5$ and the KROUPA01 IMF, while other models are designed by changing only one parameter each time
to test its effect. Fig.~1 shows the simulated cumulative XLF and its detailed components
contributed by accreting NS/BH with hydrogen-rich (NS/BH-H) and helium-rich (NS/BH-He)
MS/SG donors, and Be-XRBs (left panel) and accretion modes of simulated XRBs (right panel), respectively.
Note that our simulated XLF can match the observed average XLF pretty well.
One can see that BH-H systems dominate the XLF of both the very high luminosity
($L_{\rm X}> \sim 2\times 10^{38} \rm ergs\,s^{-1}$) and low luminosity
($L_{\rm X}<  10^{36} \rm ergs\,s^{-1}$) end, while NS-H systems play a major role
in the luminosity range of $\sim 10^{37}- \sim 10^{38} \rm ergs\,s^{-1}$. 
Moreover they are mainly persistent sources (the transients are very rare).
Our calculation shows that the BH-H ULX systems are contributed mainly by two species.
They are all persistent sources, the majority of which are mainly wind-fed
BH-XRBs with massive ($\sim 10-30 M_{\odot}$) SG donors (i.e., BH-SG HMXBs),
whose orbital period is in the range of several thousands days to even hundreds of years,
with a nearly flat eccentricity distributed from 0 to 1. The other specy is mainly RLOF-fed
BH-XRBs, with less massive (typically $< 10 M_{\odot}$) MS donors, whose orbital period is much
shorter, typically on the order of days. While BH XRBs at the low luminosity end are mainly wind-fed
BH systems powered by higher mass ($\sim 30 -75 M_{\odot}$) MS stars (i.e., BH-MS HMXBs), with
orbital period from about months to $\sim10^3$ days, as shown in Fig.~2 for the current
orbital period $P_{\rm orb}-L_{\rm X}$ (left) and $P_{\rm orb}-M_2$ (right) distribution, respectively.
In addition, the Be-XRB population is predicted to be very small. It is mainly due to the low
duty cycle transient characters of Be-XRBs relative to the long-term average of observed XLF,
supporting the expectation by \citet{bg12}.

We note that, quantitatively, our calculation is in general consistent with current
HMXB population statistics. Our prediction that XRBs with luminosity larger than
$\sim 10^{37} \rm ergs\,s^{-1}$ are mainly NS systems is in general consistent with
observational statistics by \citet{lph06}. The most luminous sources (for example ULXs)
are predicted to be BH systems, which is also not in contradiction with current observation
and theoretical expectations. The prediction that
HMXB with naked He donor stars is relatively less than HMXB containing H-rich
donors is also not at odds with current observational statistics (one He-HMXB confirmed in
the Galaxy, i.e., Cyg X-3, van Kerkwijk et al. 1992, and another two extragalactic He-HMXBs
IC10 X-1 and NGC300 X-1 confirmed by Crowther et al. 2003, 2010).
The predicted ULX HMXBs usually have massive ($\sim 10-30 M_{\odot}$) donor stars, accreted
by BHs in its SG phase, or less massive ($< \sim10 M_{\odot}$) MS donors accreted by BH in RLOF phase,
which are very similar to the sources identified by
\citet[][i.e., NGC 3031 X-11]{liu02}, \citet{robert01} and \citet[][i.e., ULX in NGC 5204]{liu04},
\citet[][i.e., NGC 1313 X-2]{zampieri04}, and \citet[][i.e., NGC 4559 X-7]{soria05}.
We also predict a preponderance of wind-fed BH HMXBs powered by massive MS stars in
relative low luminosities ($L_{\rm X}< \sim 10^{36} \rm ergs\,s^{-1}$) which has not yet
been uncovered in nearby star-forming galaxies. Future high resolution X-ray and
optical observations of this population may be used as a further test of the results obtained here.

To illustrate the formation and evolution of these BH HMXBs in detail, we
present two example evolutionary sequences for $M_1$, $M_2$, $P_{\rm
orb}$, $L_{\rm X}$ of BH-SG and BH-MS HMXBs in Figures~3 and~4, respectively. In Fig.~3, we
consider a primordial binary system in a $2505.4\,R_{\odot}$ circular
orbit. The initial stellar masses are 30.119 and 27.158$\,M_{\odot}$
for the primary and secondary, respectively. The primary first evolves across
the HG, expands and fills its RL in the core helium burning (CHeB) stage
(time 6.3257$\,$Myr), then transfers mass to the secondary star, which is still on the MS.
As mass transfer proceeds the orbit of the system first shrinks slightly then expands to
$\sim 2995.244\,R_{\odot}$ as the mass ratio gets flipped at the time of 6.5931$\,$Myr
until the binary gets detached again (time 6.9195$\,$Myr).  As the binary evolves
the orbit of the system expands slightly until the time of 7.0079 Myr when the primary forms a BH
(the CO core mass $M_{\rm CO}=7.9972M_{\odot}$, hence partial fall-back with reduced natal kicks), the
orbital separation of which is sharply increased to $\sim 5340.816\,R_{\odot}$ with a large eccentricity
of $\sim 0.46$. At this time, the system consists of a 6.50 $\,M_{\odot}$ BH and a massive
($M_2=37.181\,M_{\odot}$) MS companion. Then the rejuvenated MS star, as the primary, evolves across the HG,
expanding its radius to become a SG star at the time of $\sim 7.78\,$Myr. At this time
the SG donor which has extremely large radii ($> 1000\,R_{\odot}$) has sufficiently strong stellar winds to power
a bright HMXB activity before SN explosion which results in another BH and
the disruption of the binary system.

Next, we use a binary with initially
$M_1=39.816\,M_{\odot}$, $M_2=22.178\,M_{\odot}$ and $a=62.743\,R_{\odot}$, to give a quick
illustration of BH-MS HMXB formation and evolution, as shown in Fig.~4.
The primary fills its RL on its MS, and mass transfer proceeds as it evolves across HG till
the end of CHeB, at which point (time 5.1369$\,$Myr) it is a 12.632$\,M_{\odot}$ naked
HeMS star in a $a=124.107\,R_{\odot}$ orbit with a 45.730$\,M_{\odot}$ MS companion.
Then the HeMS star evolves across the HeHG, and explodes at the
time of 5.7680$\,$Myr, leaving a 5.497$\,M_{\odot}$ BH with a 45.327$\,M_{\odot}$ MS companion in an orbit
of 163.7139$\,R_{\odot}$. The strong stellar wind from the MS star is then accreted by the BH at a
moderately low rate compared with the SG case above, resulting in relatively low luminosity
BH-MS XRBs powered by stellar wind. As the MS star evolves across the HG and fills its RL,
the BH will spiral into its envelope due to the extremely large mass ratio, leading to a coalescence finally
at the time of 8.4605$\,$Myr, immediately after entering the CE phase.

We note that the above evolutionary sequence example may explain why BH binaries with 
massive MS donors can dominate at lower luminosities, rather than binaries with NS accretors. 
It is mainly because of the fact that the SN kicks 
the compact stars receive during SN explosion are quite different.  As illustrated above, 
the BH in BH-MS HMXB always receives a small or no SN kick, which facilitates the survival of a 
wide binary, which would probably produce a faint HMXB. However it is not the case for NSs. Due to the 
much larger SN kicks, the NS is more likely to escape from its companion, leading to the 
disruption of the binary system. Or if survived luckily it may expand its orbit greatly, showing likely
as Be/X-ray transients when active. However even so, such a channel is still insignificant when compared with 
BH-MS HMXBs, as already estimated.

In Fig.~5, we show the evolution of XLF in our basic model, in order to study the nature of
the sources, as well as its evolution. Both constant star formation (left) and a $\delta$-function 
like star formation (right) cases are studied. We note that most of the sources are produced within $\sim$ 20 Myr
after the star formation, which is in general agreement with observations \citep{sts09} and previous
studies \citep{Linden10}. We suggest that the short formation and lived time scale of sources may explain
why the universal XLF should exist naturally in the star-forming galaxies. Additionally we see that
the more recent star formation seems to have more luminous sources, resulting in a much flatter XLF at
the high luminosity end. Our results are consistent with earlier observations
\citep[see][and references therein]{fabbiano06}, revealing that the different or more complex XLFs
are mainly because of the  complexity and evolution of the X-ray source populations.
Moreover the BH-MS HMXBs seem to emerge a little earlier than BH-SG HMXBs.
This can be understood when considering the fact that the sources in both luminosity
extremes have distinct formation channels. As illustrated above, we can see that the appearance of
BH-MS sources always accompanies with the birth of the BH accretors, the progenitors of which have a
shorter nuclear evolution timescale when compared to the BH-SG HMXBs, the occurrence of which while is mainly
driven by the expanding of the less massive donor stars. However the SG donors have always much stronger
stellar winds than MS stars, leading to much brighter BH-SG HMXBs when compared to BH-MS sources.

\subsection{Effects of parameters on XLFs}
Fig.~6 shows XLFs for different models compared to our basic model M1 (solid line). Each
model is chosen to examine the effect that each parameter has on both the shape and the
absolute source number of the XLF. Several parameters have significant effects on either
the shape or the source number or both of the XLFs, while others have only minor effects.

The parameters that have minor effects include the CE efficiency parameter ($\alpha_{\rm CE}$),
 and the initial mass ratio distribution of the secondary star, as shown by models M2 and M5, respectively.
The parameter $\alpha_{\rm CE}$ dictated how efficiently orbital energy is transformed into
the kinetic energy that expels the donor's envelope during the CE phase. It mainly affects
the formation and evolution of binary systems which must go through a CE phase, such as
low mass X-ray binaries (LMXBs) \citep{Tremmel12} and cataclysmic variables \citep{paczynski76}. However this  
is not the case for HMXBs, as the major formation channels of HMXBs do not involve 
CE phases as severely as LMXBs \citep{Linden10,Valsecchi10}.
We also change $\alpha_{\rm CE}$ to other values and forms \citep[for details see][and comparisons with the $\gamma$-algorithm]{zuo13}.
Changing the initial binary mass ratio from a flat distribution (model M1)
to a ``twins" distribution (model M5) has little effect on the XLF, although there is a
slight increase in the number of bright HMXBs. This is because HMXBs require mass ratios
close to one which is achieved by the ``twins" model which forces mass ratios close to unity.

The binary fraction only affects the absolute source number of the XLF. As shown in Fig.~5
an increase of binary fraction (i.e., model M3) means more XRBs are
produced, hence an overall shift of the XLF curve compared to that of the basic model.
A flatter IMF (i.e., model M6) implies a larger
number of massive stars, resulting in more compact objects compared to a steeper one.
Hence a flatter IMF will results in more luminous HMXBs.

We suggest that the diversity of stellar components may explain the normalization dispersions of XLFs
between galaxies.  We note that the simulated star-forming galaxy in our basic model only represents a typical
case for this kind of galaxies. While for each individual galaxy, it may have its specific stellar
properties, such as different stellar mass distributions (for both the primary and the secondary), and different binary
fractions. Our parameter studies (IMF, f and $P(q)$) precisely support the idea that the normalization dispersion
is of a physical origin, proposed by MGS. However we emphasize that
the intrinsic physics governing the binary evolution should keep the same for binaries in these galaxies, as examined below.

In model M4, a significant luminosity break emerges when decreasing the 'Begelman factor'
$\eta_{\rm Edd}$ by a factor of $\sim$20 (dotted line in Figures~6 and 7, respectively).
A similar trend has been found previously \citep{liu07,Linden10}.
In order to better constrain the super-Eddington factor, we
modify the 'Begelman factor' to 80 (dash-dotted line), 50 (dash-dot-dotted line),
30 (long-dashed line), 10 (short-dashed line), respectively, as shown in Fig.~7. We note the luminosity break
exists clearly even for $\eta_{\rm Edd}$ as high as $\sim30-50$. This marked contrast with the observed smooth XLF
implies that the actual maximum luminosities of accreting BHs can be as high as even $\sim 100$
times the corresponding Eddington luminosities, as suggested by \citet{Begelman02}.

Increasing the dispersion velocity $\sigma_{\rm kick}$ means that natal kicks of high
magnitude are chosen more frequently from the Maxwellian distribution. As a consequence more
binaries are disrupted during the SN explosions. This decreases the pool of potential HMXBs and
may account for the smaller number of HMXBs in model M7 (dash-dot-dotted line) as shown in Figures~6 and 8, respectively.
Additionally, a larger natal kick can move the wind-fed BH-MS HMXBs into
a much wider orbit, too widely separated for stellar material to be effectively accreted onto the BH,
 resulting in much lower luminosities, and hence a smaller number of BH-MS XRBs in this
luminosity range. This phenomenon is shown clearly in Fig.~8 where the dispersion of kick velocity $\sigma_{\rm kick}$ is
modified to 190 (dash-dotted line), 170 (dotted line), 100 (short-dashed line), and 50 (long-dashed line), respectively. We note that 
compared with our basic model the predicted number of low luminosity HMXBs decreases with increasing
natal kicks, while smaller $\sigma_{\rm kick}$ may increase the formation rate of HMXBs remarkably.
Specifically, models with high natal kicks (i.e., model M7) predict significantly less HMXBs than
are observed, while models with low natal kicks (i.e., $\sigma_{\rm kick}<\sim100\,\rm km\,s^{-1}$) predict
too many HMXBs. Based on these results, we conclude that our models in which
the natal kick velocity dispersion above $\sim200\,\rm km\,s^{-1}$ or below $\sim100\,\rm km\,s^{-1}$
are inconsistent with the observations. The typical kicks that match
the observed HMXB XLF are on the order of $\sigma_{\rm kick}\sim150\,\rm km\,s^{-1}$.
Using a similar method \citet{bel10b} proposed a comparable value of natal kick dispersion
($\sigma_{\rm kick}\sim170\,\rm km\,s^{-1}$ ) to match the observed intrinsic ratio of
double and single recycled pulsars in the Galactic disc. Our finding is in
general consistent with theirs. It is not surprising as isolated recycled pulsars
and double neutron star (DNS) binaries are both presumably the descendant of NS HMXBs.
Our conclusions may further support earlier findings that NSs formed in binaries receive significantly
smaller natal kicks than the velocities of Galactic isolated pulsars would seem to indicate.

Stellar winds play an important role in the evolution of high mass stars in two
major competing ways. A stronger stellar wind will increase the accretion rate of
wind-fed HMXBs, making it more luminous. On the contrary, a weaker stellar wind will
result in a larger pre-SN mass, and hence the formation of more numerous and
more massive BHs. This may increase the luminosities of HMXB populations, as on the
one hand, BH-XRBs can form stable RLO XRBs with more massive companions compared to
NS-XRBs, and on the other hand, more massive BHs may drive higher accretion rate, and
therefore higher luminosities. Comparing models M1 and M8, we can see that weaker
stellar winds increase both the number and luminosity of bright HMXBs, so the latter
effect is the dominant one. We note here that our findings are also consistent with the 
results obtained by \citet{f13} and \citet{Tremmel12}.

Our results are subject to some uncertainties and simplified treatments.
For example, in our calculations, 
only HMXBs with stellar mass BHs are considered. However IMBH which is presumably formed through BH
mergers may also show up as ULXs. Though it is expected to be  significantly less frequent
than stellar mass BHs, we should caution that, even only one of
this kind of source may change the high luminosity tail of XLF significantly.
A further careful modeling of IMBH considering dynamical formation processes
may resolve this problem, however it is beyond the scope of this paper.
On the other hand in the framework of stellar mass BHs, we may see that
the ULX population can be generally accounted for by normal HMXBs, only in the
case of mild super-Eddington accretion rate allowed. Additionally, since little is yet known about, either the detail SFH and IMF
in star-forming galaxies, or key processes, such as the detailed
accretion modes in XRBs, it is difficult to ascertain which
parameter combinations are the best or most realistic by comparison with observations.
For example, the normalization of the simulated XLF depends on the adopted values of
several parameters, such as the bolometric correction factor $\eta_{\rm bol}$ and binary fraction $f$.
These two parameters show some degeneracy, and a slightly lower bolometric correction factor would
favor a larger binary fraction.
However the overall shape of the simulated XLF depends most strongly on two parameters:
the natal kick dispersion $\sigma_{\rm kick}$ and the allowed boost factor of super-Eddington
accretion rate $\eta_{\rm Edd}$. The former, related to the binary interactions, determines
the final outcome of the SN explosion. The latter is related to the accretion
behavior, and constrains the location of the break in the XLF. They jointly determine
the shape of the XLF. Conversely, the confirmed universal featureless XLF can make a
good decision for the precise choice of the corresponding parameters.

\section{SUMMARY}

We have used an EPS code to model the universal featureless XLF of HMXBs
in star-forming galaxies. We used the apparent universal XLF to constrain
models of XRBs. Our study shows that the single, smooth power law XLF can
be excellently reproduced with all models considered, but with two parameters
strongly affecting its overall shape: the dispersion of natal kick velocity
$\sigma_{\rm kick}$ and the introduced parameter ``Begelman factor" $\eta_{\rm Edd}$.
The overall shape and normalization of HMXB XLF need the natal kick
dispersion $\sigma_{\rm kick}\sim150\,\rm km\,s^{-1}$, which is
generally consistent with the finding by \citet{bel10b} based on the statistics of
double and single recycled pulsars. Our XLF modeling further strengthens earlier
finding that NSs formed in close interacting binaries receive significantly smaller natal kicks than
the velocities of Galactic single pulsars would indicate.
The absence of features in the XLF near the critical Eddington luminosity of a NS or a
stellar-mass BH and the cut-off luminosity at $L_{\rm X} \sim 10^{40} \rm\, ergs\,s^{-1}$
need the allowed boost factor of super-Eddington accretion rate as high as
$\sim 80-100$. Our results give strong supports for the suggestion by \citet{ggs03}
that the bulk of ULXs may indeed be the high-luminosity extension of
ordinary HMXBs which harbor stellar-mass BHs with mildly super-Eddington accretion
rate, rather than exotic intermediate-mass objects.
We present the detail components
of HMXB populations which contribute to the observed XLF, and emphasize that 
the low luminosity sources of $L_{\rm X} <10^{36} \rm\, ergs\,s^{-1}$ are mainly
wind-fed BH systems powered by high mass ($\sim30-75\,M_{\odot}$) MS stars with
orbital periods around months to $\sim10^3$ days
which have not yet been verified in nearby star-forming galaxies
due to limited instrument capabilities. Our work motivates
further high-resolution X-ray and optical observations of HMXB populations
in nearby star-forming galaxies.

\acknowledgments We thank Marc van der Sluys for providing routines to
compute envelope binding energies of giant stars and helpful discussions. 
We thank Zhi-Yuan Li for his assistance with the language improvement. 
This work was supported by the National Natural Science
Foundation (grants 11103014, 11133001, 10873008 and 11003005), the Research Fund for
the Doctoral Program of Higher Education of China (under grant number 20110201120034), the National
Basic Research Program of China (973 Program 2009CB824800), the Fundamental Research Funds for the
Central Universities and National High Performance Computing Center (Xi'an).

\newpage

\begin{table}
\caption{Parameters adopted for each model. Here $\alpha_{\rm CE}$
is the CE parameter, $q$  the initial mass ratio, IMF is the
initial mass function, $f$ binary fraction, $\eta_{\rm Edd}$ - the factor of super-Eddington
accretion rate allowed, $\sigma_{\rm kick}$ the dispersion of kick speed,
STD is the standard stellar winds while WEAK represents the standard wind mass loss
rate reduced to 50\%.
} \centering
\begin{tabular}{ccccccccc}\hline\hline
   Model & $\alpha_{\rm CE}$ & P(q)    & IMF &  $f$  &  $\eta_{\rm Edd}$   & $\sigma_{\rm kick}$ & winds\\ \hline
      M1 & 0.5 & $\propto q^{0}$ & KROUPA01     &   0.5 & 100  & 150 & STD  \\
      M2 & 1.0 & $\propto q^{0}$ & KROUPA01     &   0.5 & 100  & 150 & STD  \\
      M3 & 0.5 & $\propto q^{0}$ & KROUPA01     &   0.8 & 100  & 150 & STD  \\
      M4 & 0.5 & $\propto q^{0}$ & KROUPA01     &   0.5 & 5   & 150 & STD  \\
      M5 & 0.5 & $\propto q^{1}$ & KROUPA01     &   0.5 & 100  & 150 & STD  \\
      M6 & 0.5 & $\propto q^{0}$ & MT87         &   0.5 & 100  & 150 & STD  \\
      M7 & 0.5 & $\propto q^{0}$ & KROUPA01     &   0.5 & 100  & 265  & STD  \\
      M8 & 0.5 & $\propto q^{0}$ & KROUPA01     &   0.5 & 100  & 150 & WEAK  \\  \hline
\end{tabular}
\end{table}

\begin{figure}
  \centering
   \includegraphics[width=0.48\linewidth]{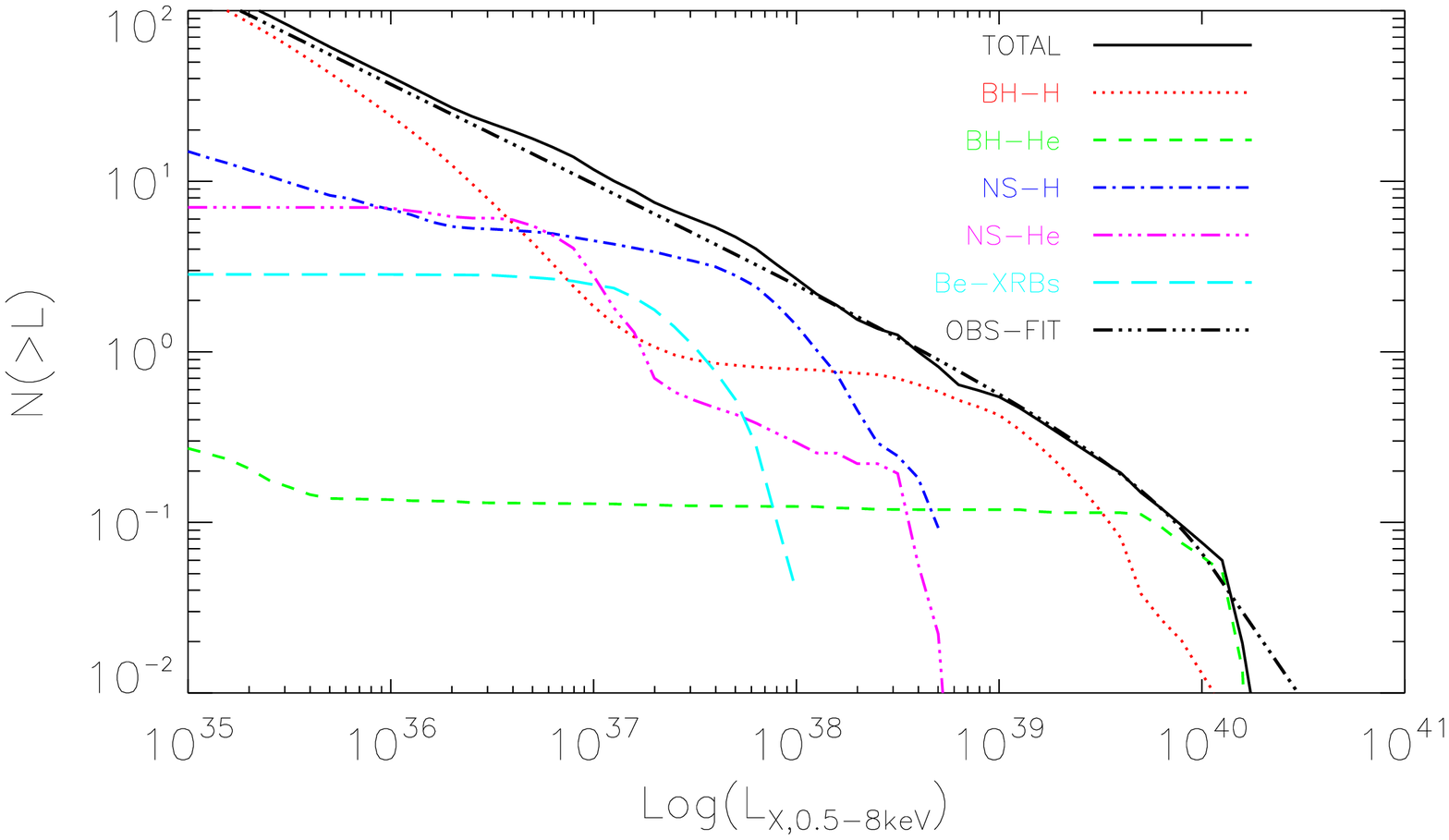}
      \includegraphics[width=0.48\linewidth]{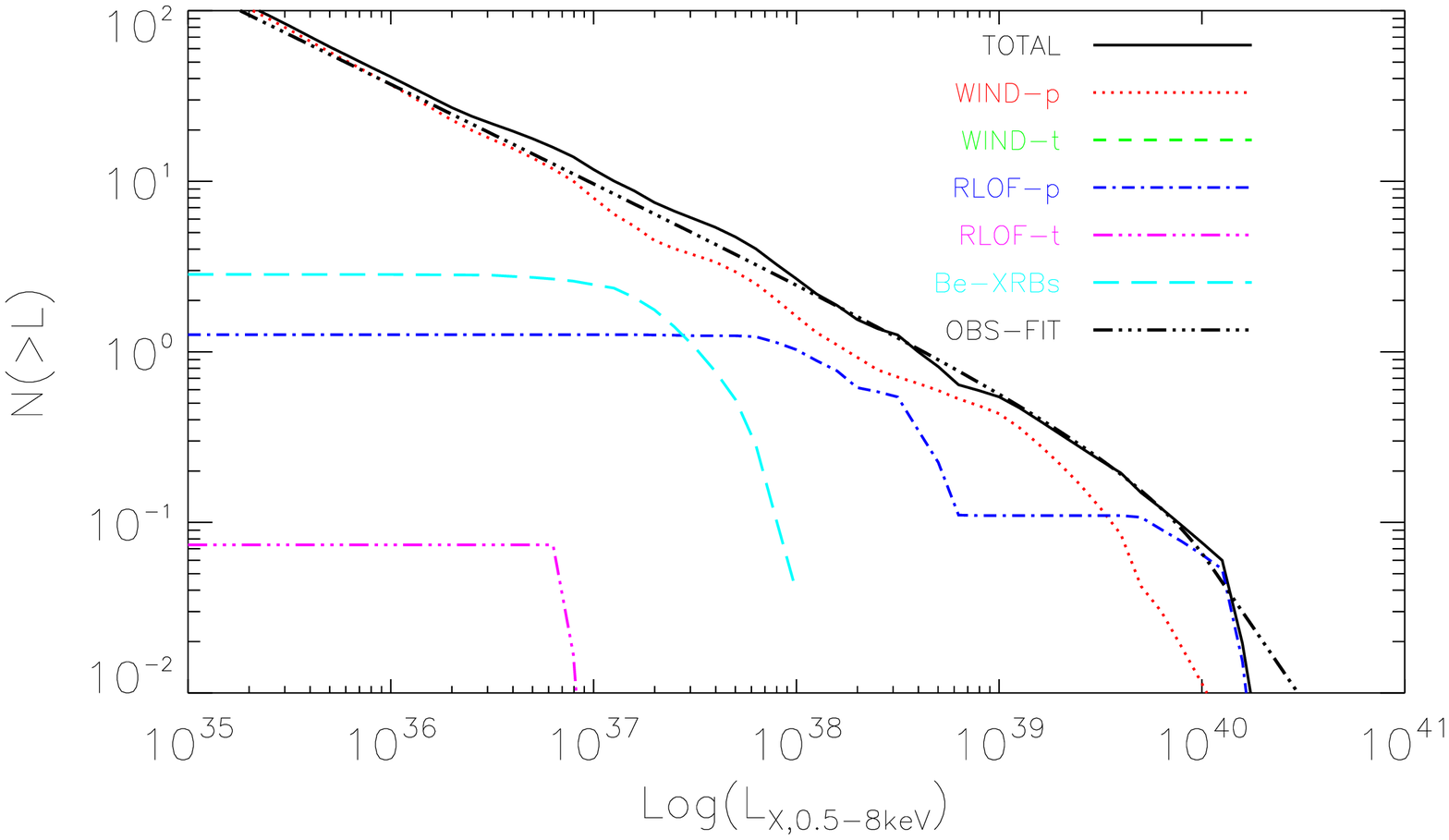}
\caption{The detailed components of the simulated XLF (\emph{left})
and accretion modes of simulated XRBs (\emph{right}) in model M1. \emph{Left} panel: The solid,
dotted, short-dashed, dash-dotted, dash-dot-dotted, long-dashed lines represent ALL-XRBs,
BH-H, BH-He, NS-H, NS-He MS/SGXRBs and Be-XRBs, respectively. \emph{Right} panel: The solid,
dotted, short-dashed, dash-dotted, dash-dot-dotted, long-dashed lines represent ALL-XRBs,
wind-fed persistent (WIND-p), wind-fed transient (WIND-t), RLOF-fed persistent (RLOF-p), RLOF-fed transient (RLOF-t) sources and Be-XRBs, respectively.
The thick dash-dot-dotted line represents the derived average XLF (labeled as
``OBS-FIT") by MGS using the data of 29 nearby star-forming galaxies.}
  \label{Fig. 1a}
\end{figure}

\begin{figure}
  \centering
   \includegraphics[width=0.45\linewidth]{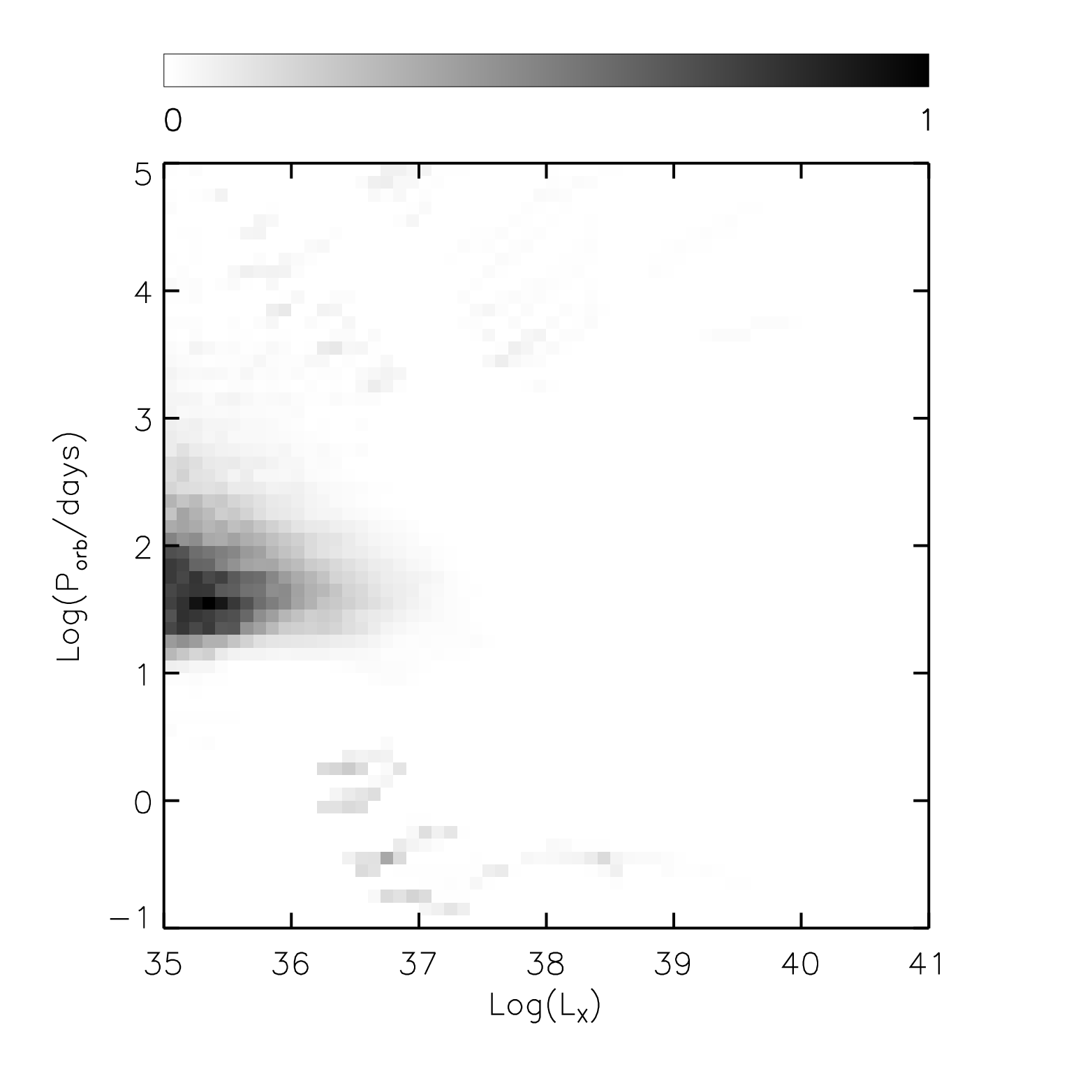}
      \includegraphics[width=0.45\linewidth]{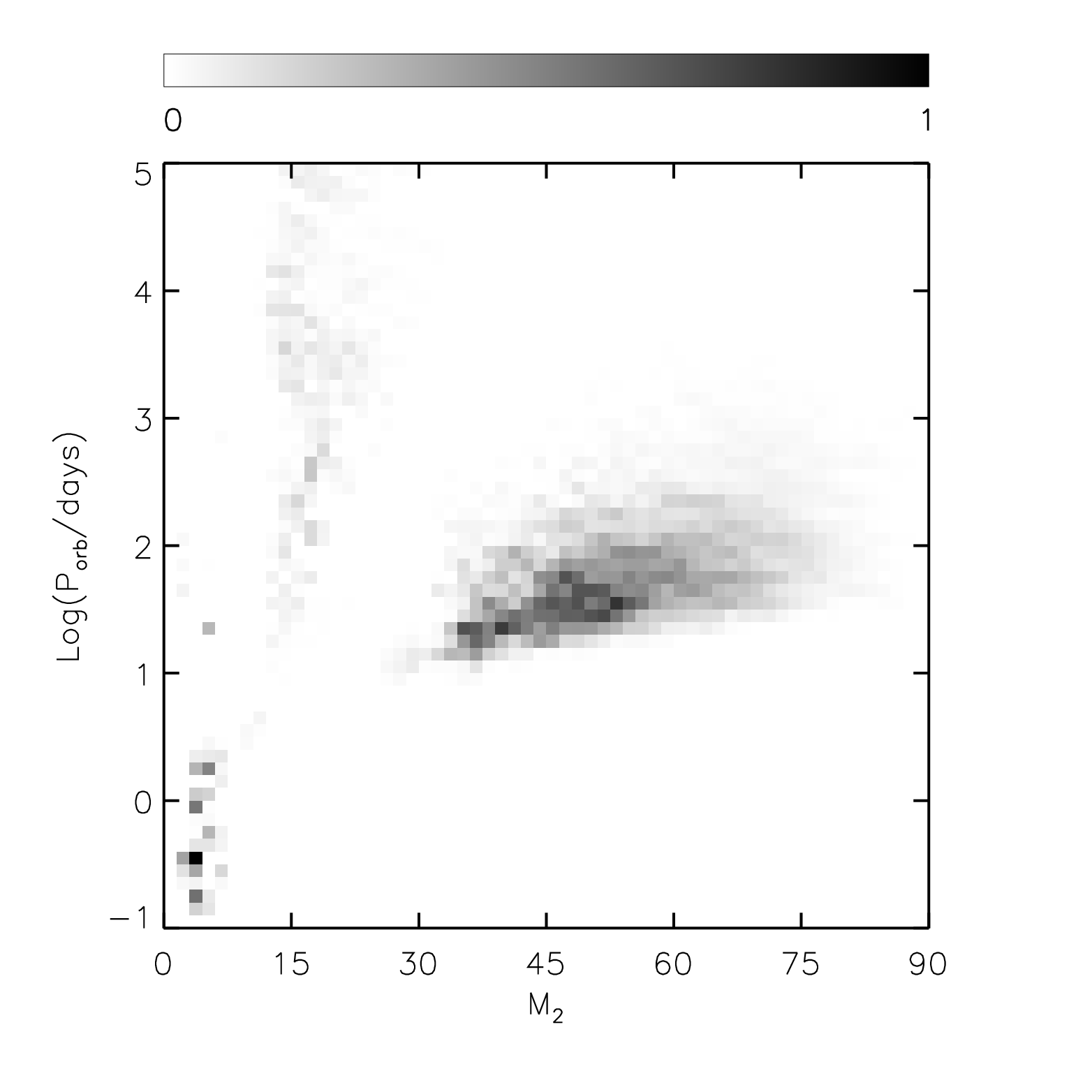}
\caption{The current orbital period $P_{\rm orb}-L_{\rm X}$ (left)
and $P_{\rm orb}-M_2$ (right) distributions in model M1.}
  \label{Fig. 1a}
\end{figure}

\begin{figure}
  \centering
   \includegraphics[width=0.8\linewidth]{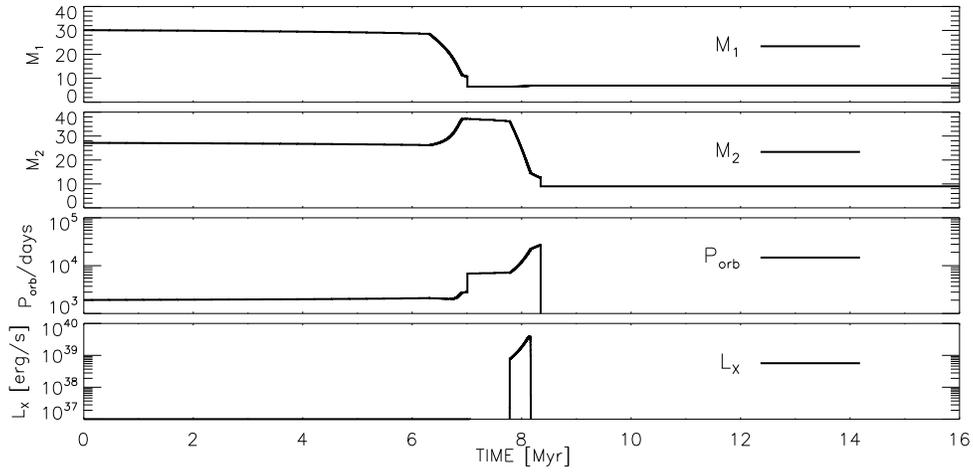}
\caption{The evolution of $M_1$, $M_2$, $P_{\rm orb}$, and $L_{\rm X}$ for
an example of bright BH-SG HMXBs in model M1.}
  \label{Fig. 1a}
\end{figure}

\begin{figure}
  \centering
   \includegraphics[width=0.8\linewidth]{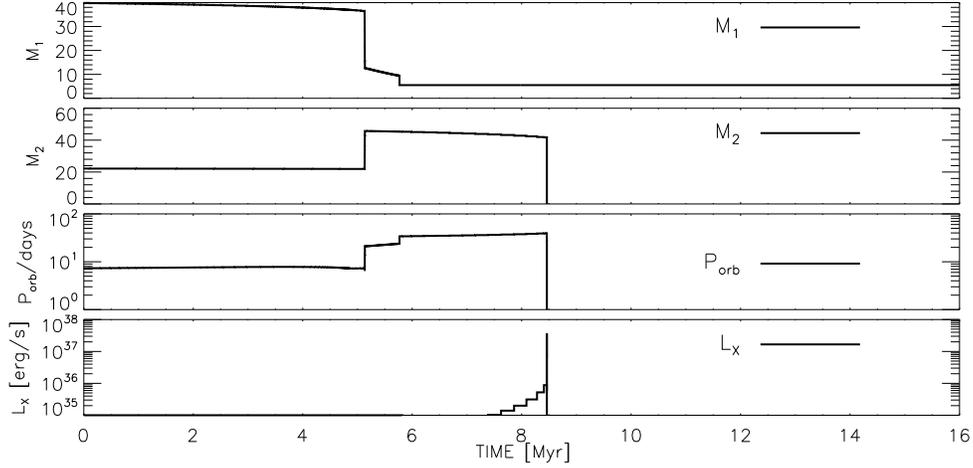}
\caption{The evolution of $M_1$, $M_2$, $P_{\rm orb}$, and $L_{\rm X}$ for
an example of low luminosity BH-MS HMXBs in model M1.}
  \label{Fig. 1a}
\end{figure}


\begin{figure}
  \centering
   \includegraphics[width=0.45\linewidth]{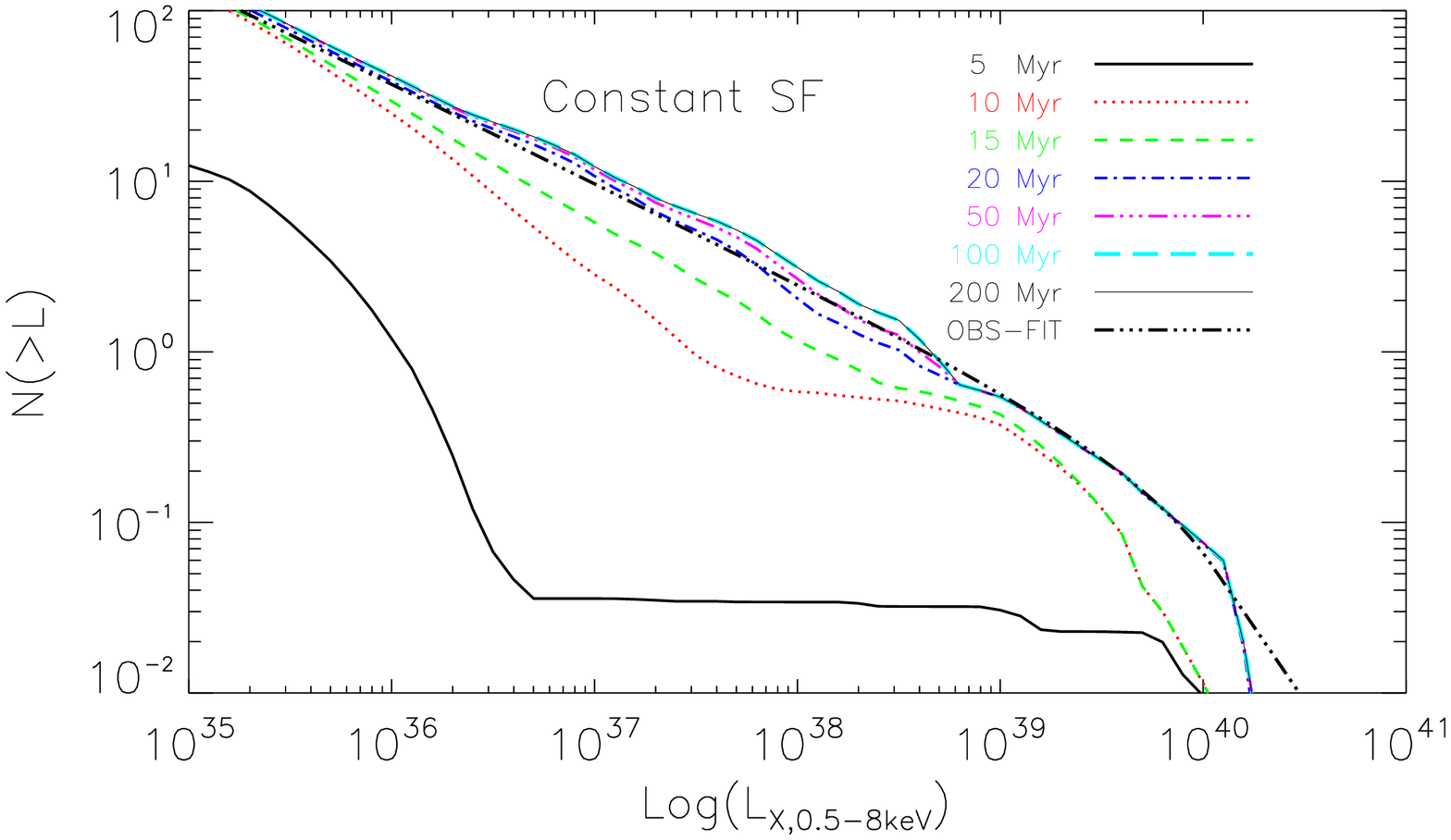}
      \includegraphics[width=0.45\linewidth]{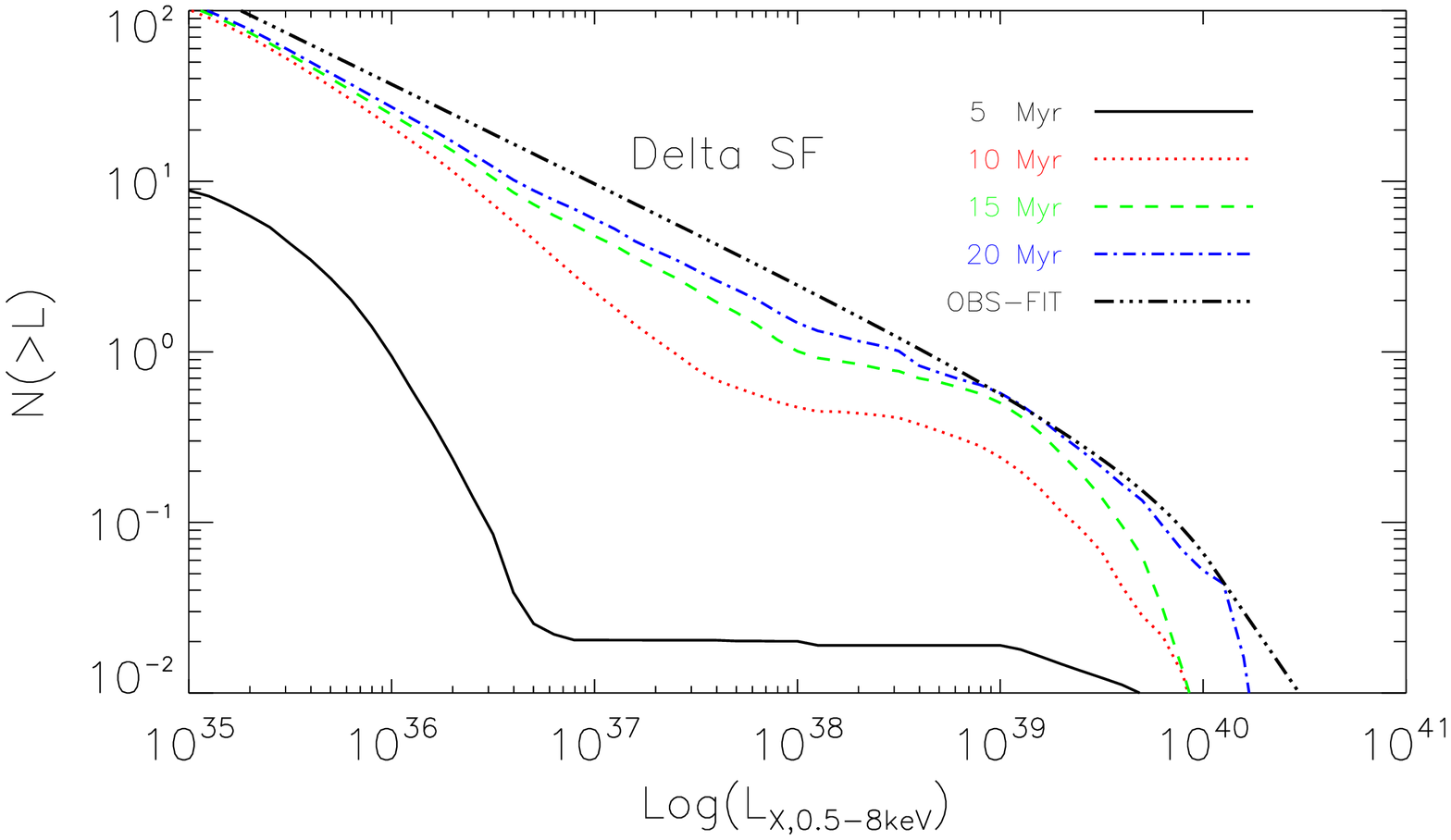}
\caption{The evolution of XLF in model M1. \emph{Left} panel: We adopted a constant star formation for 5 Myr (thick solid line), 10 Myr (dotted line), 15 Myr (short-dashed line), 20 Myr (dash-dotted line), 50 Myr (dash-dot-dotted), 100 Myr (long-dashed line), and 200 Myr (thin solid line), respectively. \emph{Right} panel: A $\delta$-function like star formation episode is adopted. Here the star formation history is set as 20 Myr, with peak $\rm SFR=1\,M_{\odot}/yr$ in the middle. The solid, dotted, dashed and dash-dotted lines represent the XLFs at the age of 5 Myr, 10 Myr, 15 Myr and 20 Myr since the beginning of the star formation, respectively. The thick dash-dot-dotted line represents the observed average XLF.}
  \label{Fig. 1a}
\end{figure}

\begin{figure}
  \centering
   \includegraphics[width=0.8\linewidth]{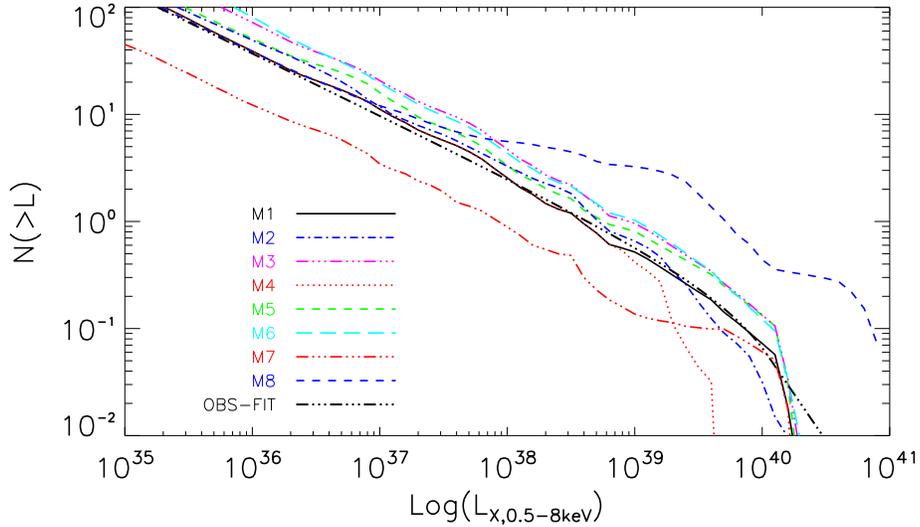}
\caption{Comparisons of the simulated and observed average (labeled as ``OBS-FIT", thick dash-dot-dotted line) XLFs.
Compared to the basic model M1 (solid line), in M2 (dash-dotted line) the CE parameter
$\alpha_{\rm CE}$ is increased to a value 1.0, in M3 (dash-dot-dotted line),
the binary fraction $f$ is set as 0.8. In M4 (dotted line), the factor
for super-Eddington accretion rate is decreased by a factor of 20. We take an atypical
distributions of mass ratio in M5 (short-dashed line) and a flatter IMF in M6 (long-dashed line),
respectively. In M7 (dash-dot-dotted line), the dispersion of kick speed is
increased to $\sigma_{\rm kick} = 265 \,\rm km\,s^{-1}$. We reduce the standard wind mass loss rate by
a factor of 2 in M8 (short-dashed line).}
  \label{Fig. 1a}
\end{figure}

\begin{figure}
  \centering
   \includegraphics[width=0.8\linewidth]{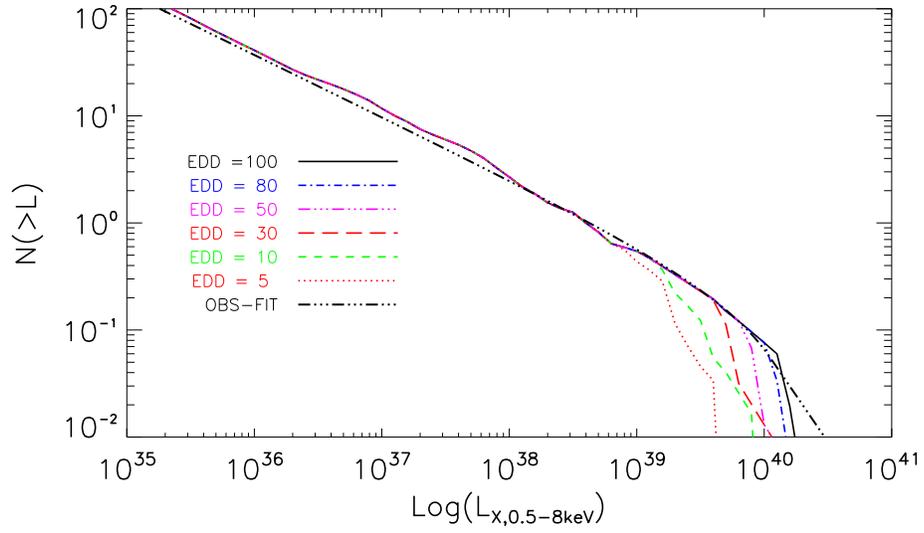}
\caption{Comparisons of the simulated XLFs with super-Eddington factors ($\eta_{\rm Edd}$)
adopted as 100 (the basic model, M1, solid line), 80 (dash-dotted line), 50 (dash-dot-dotted line),
30 (long-dashed line), 10 (short-dashed line), and 5 (dotted line), respectively. The thick
dash-dot-dotted line represents the observed average XLF.}
  \label{Fig. 1a}
\end{figure}

\begin{figure}
  \centering
   \includegraphics[width=0.8\linewidth]{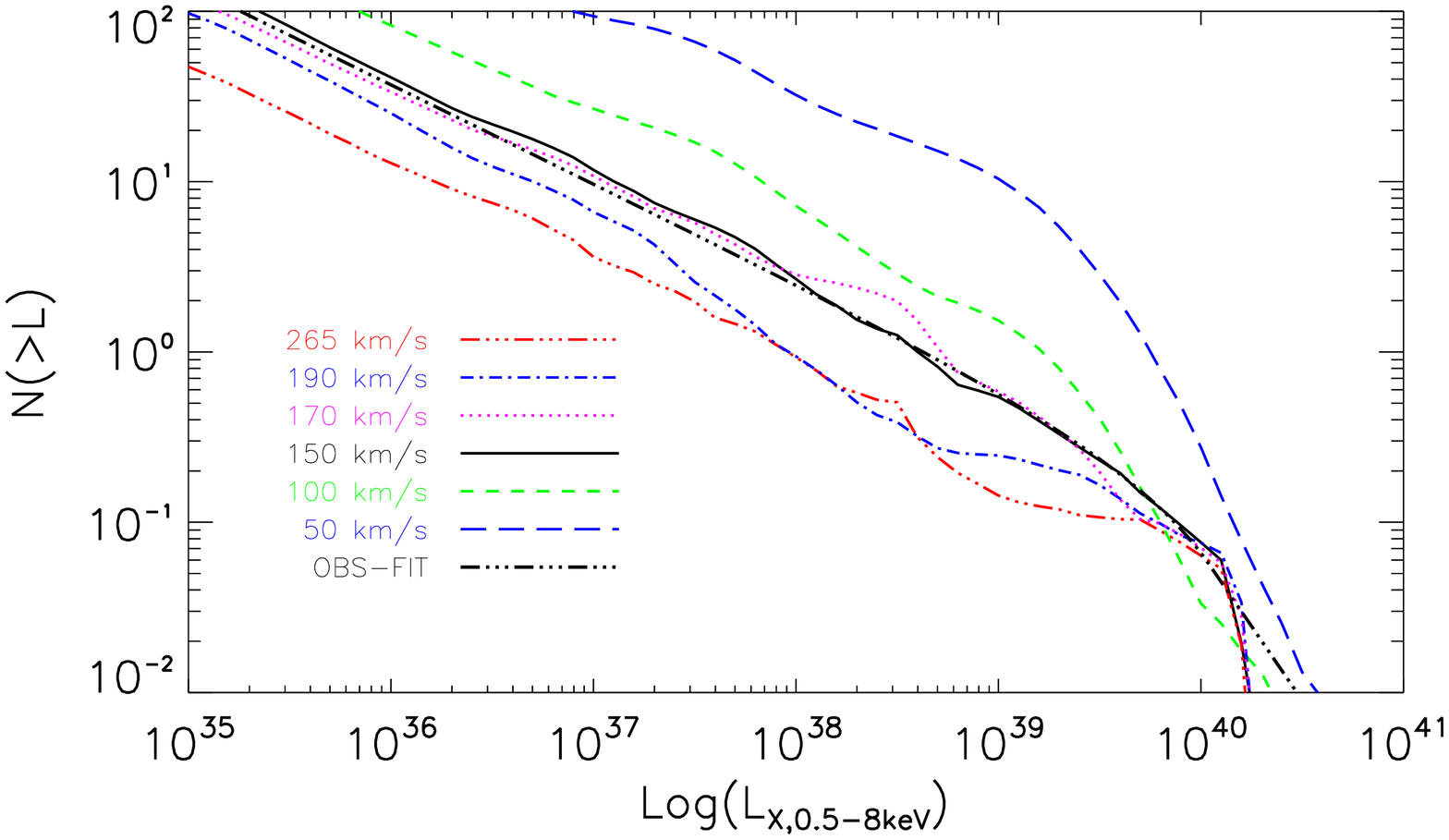}
\caption{Comparisons of the simulated XLFs with the dispersion of kick velocity ($\sigma_{\rm kick}$)
adopted as 265 (dash-dot-dotted line), 190 (dash-dotted line), 170 (dotted line),
150 (the basic model, M1, solid line), 100 (short-dashed line) and 50 (long-dashed line), respectively. The thick
dash-dot-dotted line represents the observed average XLF.}
  \label{Fig. 1a}
\end{figure}

\newpage
\section{APPENDIX A}

Rather than a constant value adopted conventionally, a critical mass ratio $q_{\rm cr}$
determining the allowed parameter space for stable mass transfer in the $P_{\rm orb}-M_1$
plane is developed recently by \citet{shao12}. For a specific binary consisting of a
massive primary star and a less massive secondary, if $P_{\rm orb}$ is initially too short,
the orbital separation will always decrease with mass transfer, a sufficiently dense gas
flow may exceed the Roche lobe, leading to a CE phase. On the other hand, if the
$P_{\rm orb}$ is too long, the primary may have climbed to the (super)giant branch and
developed a deep convective envelope around the compact core prior to mass exchange,
a runaway mass transfer will happen, leading to the CE evolution. Thus for each mass ratio
$q$, there exist both upper and lower limits of the orbital period ($P_{\rm orb,up}$
and $P_{\rm orb,low}$), between which the binary can evolve smoothly with stable mass
transfer on thermal timescale.

\citet{shao12} provide two choices of metallicity ($Z=0.001$ and $Z=0.02$). In our cases,
the higher and more appropriate value, i.e., $Z=0.02$ is adopted, the grid of which we believe can be used
without too much loss in accuracy. The corresponding upper and lower limits of orbital period ($P_{\rm orb,up}$ and $P_{\rm orb,low}$)
for a certain mass ratio $q$ are fitted as a function of initial primary mass ($M_1$)
in the form of binomial: $$P_{\rm orb}=\sum_{n=0}^{5} a_{\rm n}M_1^{\rm n}$$ 
coefficients of which are listed in Table~2. The upper part of Table~2 is for lower limits,
with eleven discrete values of $q$ in the range of 2 to 12. The upper limits of orbital
period for each $q$ are very similar, so we give only one rough fitting, the coefficients of
which are listed in the lower part of Table~2 (labeled as 'ALL').

\begin{table}
\caption{Fitting formula coefficients of CE criterion. Here $q$ is
the initial mass ratio, $a_{\rm n}$ (n=0-5) is the coefficient of the binomial.
Note that a blank entry in the table implies a zero value.} \centering
\begin{tabular}{cccccccc}\hline\hline
    q  &     $a_0$    &    $a_1$   &     $a_2$   &    $a_3$    &     $a_4$      & $a_5$   \\ \hline
    2  &    -7.55E-1  &  4.577E-1  &    -2.06E-2 &   3.88E-4   &     -2.57E-6   &   \\
  2.5  &    -1.955E+0  &  1.026E+0   &   -4.937E-2 &   9.36E-4   &     -6.14E-6   &  \\
    3  &    -3.278E+0  &  1.565E+0   &   -6.628E-2 &   1.07E-3   &     -6.01E-6   &  \\
  3.5  &    2.4E-1    &  4.05E-1   &    1.299E-1 &   6.44E-3   &     1.06E-4    & -5.75E-7  \\
    4  &    -1.341E+1 &  5.232E+0   &    -1.496E-1&   1.144E-3  &                &  \\
  4.5  &    1.143E+2  &  -1.997E+0  &    4.168E-3 &             &                &  \\
    5  &    1.612E+2  &  -2.44E+0   &    -6.85E-4 &             &                &  \\
    6  &    1.911E+2  &  1.558E+1  &    -6.27E-1 &  5.35E+3    &                &  \\
    8  &    5.8938E+3 &  -1.9817E+2&    1.6907E+0 &             &                &  \\
    10 &    3.96712E+4& -1.38555E+3&    1.2168E+1&             &                &  \\
    12 &    7.86627E+4& -2.64895E+3&    2.2478E+1&             &                &        \\ \hline
  ALL  &    8.9493E+0  & -2.7841E+3 &    6.285E-1 &   2.7E-3    &                &         \\ \hline
  \end{tabular}
\end{table}

\end{document}